\newcommand{\bscco}{BSCCO}

\documentclass[prl,twocolumn,amsmath,amssymb,showpacs,noshowkeys,superscriptaddress]{revtex4}

\usepackage{graphicx}
\usepackage{dcolumn}
\usepackage{bm}
\usepackage{color}

\begin{document}

\title{Transport Properties Governed by the Inductance of the Edges in Bi$_2$Sr$_2$CaCu$_2$O$_8$}

\author{H.~Beidenkopf}
\email{haim.beidenkopf@weizmann.ac.il}
\author{Y.~Myasoedov}
\author{E.~Zeldov}
\affiliation{Department of Condensed Matter Physics, Weizmann
Institute of Science, Rehovot 76100, Israel}%
\author{E.H.~Brandt}
\affiliation{Max-Planck-Institut f\"{u}r Metallforschung,
Heisenbergstr. 3, D-70506 Stuttgart, Germany}
\author{G.P.~Mikitik}
\affiliation{Max-Planck-Institut f\"{u}r Metallforschung,
Heisenbergstr. 3, D-70506 Stuttgart, Germany} \affiliation{B. Verkin
Institute for Low temperature Physics \& Engineering, Kharkov 61103,
Ukraine}
\author{T.~Tamegai}
\affiliation{Department of Applied Physics, The University of Tokyo, Hongo, Bunkyo-ku, Tokyo 113-8656, Japan}%
\author{T.~Sasagawa}
\affiliation{Materials and Structures Laboratory, Tokyo Institute of
Technology, Kanagawa 226-8503, Japan}
\author{C.~J.~van der Beek}
\affiliation{Laboratoire des Solides Irradi\'{e}es, CNRS UMR 7642 \&
CEA/IRAMIS/DRECAM, Ecole Polytechnique, 91128 Palaiseau cedex,
France}

\date{\today}

\begin{abstract}
We study the distribution of transport current across
superconducting Bi$_2$Sr$_2$CaCu$_2$O$_8$ crystals and the vortex
flow through the sample edges. We show that the $T_x$ transition is
of electrodynamic rather than thermodynamic nature, below which
vortex dynamics is governed by the edge inductance instead of the
resistance. This allows measurement of the resistance down to two
orders of magnitude below the transport noise. By irradiating the
current contacts the resistive step at vortex melting is shown to be
due to loss of c-axis correlations rather than breakdown of
quasi-long-range order within the a-b planes.
\end{abstract}





\maketitle

Numerous phase transitions have been proposed to interpret the intricate $B-T$
phase diagram of vortex matter in the high temperature superconductor
Bi$_2$Sr$_2$CaCu$_2$O$_8$ (BSCCO) \cite{blatter.g.rmp66,mikitik.gp.prb68}. The
first-order melting at $T_m$, that separates a quasi-ordered vortex solid from a
vortex liquid, is widely accepted to be a genuine thermodynamic phase transition
\cite{zeldov.e.nat375, schiling.a.prl78}. Experiments indicate that the glass line,
$T_g$, is another thermodynamic transition that apparently separates amorphous
solid from liquid at high fields
\cite{safar.h.prl68,vdbeek.cj.pc195,beidenkopf.h.prl95,beidenkopf.h.prl98} and
Bragg glass from depinned lattice at low fields
\cite{beidenkopf.h.prl95,beidenkopf.h.prl98}. On the other hand, the $T_x$
transition, which resides above $T_m$ and $T_g$, has remained highly controversial.
A number of experiments \cite{fuchs.dt.prl80,forgan.em.czec46,
shibauchi.t.prl83,ando.y.prb59,kimura.k.pb284,eltsev.y.pc341}, numerical
simulations \cite{sugano.r.pb284,nonomura.y.prl86}, and theoretical studies
\cite{li.d.pc468,dietel.j.prb79} argued that it is a transition into a phase with
intermediate degree of order such as a disentangled liquid \cite{samoilov.av.prl76}
or a decoupled- \cite{glazman.li.prb43,horovitz.b.prl80}, soft-
\cite{carruzzo.hm.pmb77}, or super-solid \cite{frey.e.prb49,feigelman.mv.pc167}. In
this letter we show that $T_x$ does not represent a thermodynamic transformation of
a bulk vortex property, but rather reflects an \emph{electrodynamic} crossover in
the dynamic response of the sample \emph{edges}. The inductance of the sample
edges, though immeasurably low, completely governs the vortex dynamics below $T_x$.
We use this finding to investigate the resistance due to vortex motion down to two
orders of magnitude below the sensitivity of transport measurements.

\begin{figure} [!b] \centering
\mbox{\includegraphics[width=0.49\textwidth]{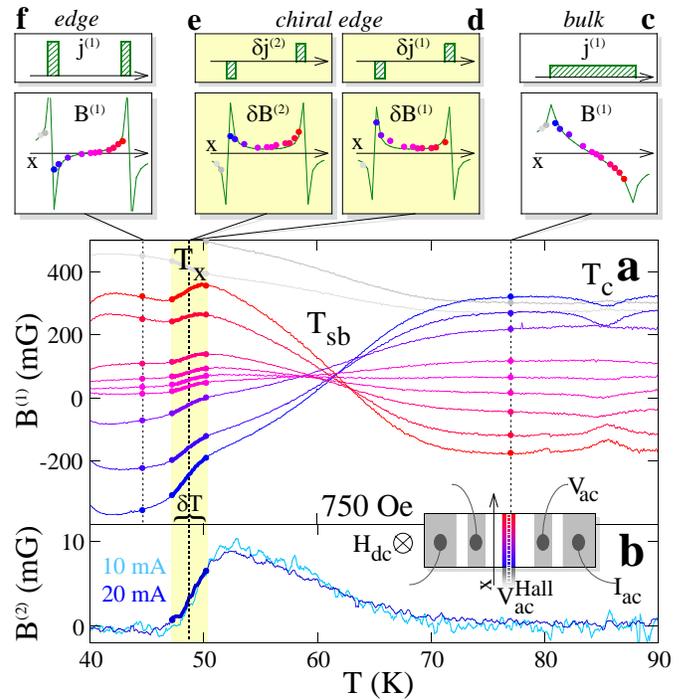}}
\caption{(color online) Hall sensor array measurement of the ac
magnetic field self-induced by a 15 mA, 73 Hz ac current in \bscco{}
crystal A. (a) First-harmonics of individual sensors. (b)
Second-harmonics from a sensor close to the edge at two current
amplitudes normalized to 10 mA. (c-f) Magnetic profiles measured
($\bullet$) and calculated (solid lines) from current profiles
(upper panels) using the Biot-Savart law: (c) Uniform bulk flow at
high temperatures turns to (f) edge flow at low temperatures. Chiral
edge flow vanishes at $T_x$ from (d) first- and (e)
second-harmonics. Inset in (b): Schematic location of Hall sensors
relative to the contacted sample.} \label{fig2}
\end{figure}

The resistance of several \bscco{} single crystals of typical size
1500$\times$350$\times$20 $\mathrm{\mu m^3}$ in a dc magnetic field, $H_{dc}
\|$c-axis, was measured using two complimentary techniques: directly via transport
and indirectly by determining the ac current distribution with Biot-Savart law from
the self-induced ac magnetic field profile using an array of 10$\times$10
$\mu$m$^2$ GaAs Hall sensors. In addition, to eliminate the c-axis contribution to
the measured resistance we irradiated the current contacts at GANIL (Caen, France)
with 1 GeV Pb ions to a matching dose of $B_\Phi=0.5$ T, while masking the rest of
the sample, which remained pristine.

Figure \ref{fig2}(a) shows the temperature dependence of the current-induced ac
field, $B^{(1)}(x)$. At high temperatures the current distributes uniformly because
it is solely dictated by the flux flow resistance in the bulk, $R_b(T)$.
Accordingly, $B^{(1)}(x)$ measured by the Hall sensors across the sample
(Fig.~\ref{fig2}(c), circles) fits perfectly to that calculated via Biot-Savart law
(solid line) from a uniform current $j^{(1)}(x)$ (upper panel).

With cooling $B^{(1)}(x)$ gradually flattens out, becomes completely
flat at $T_{sb}$, and eventually inverted at yet lower temperatures.
The inverted $B^{(1)}(x)$ profile is associated with essentially
pure edge currents (Fig.~\ref{fig2}(f)). The Bean-Livingston surface
barrier \cite{bean.cp.prl12} and the platelet sample geometry
\cite{zeldov.e.prl73} impose an energetic barrier that progressively
impedes vortex passage through the edges. This defines an effective
edge resistance, $R_e(T)$, that decreases rapidly with cooling and
becomes smaller than the bulk resistance $R_b(T)$ below $T_{sb}$. As
a result, most of the current shifts from the bulk, where little
force is required to move vortices across, towards the edges, where
it facilitates the hindered vortex entry and exit
\cite{fuchs.dt.nat391}. Bulk vortex pinning becomes dominant only at
significantly lower temperature (not shown).

\begin{figure} [!t] \centering
\mbox{\includegraphics[width=0.49\textwidth]{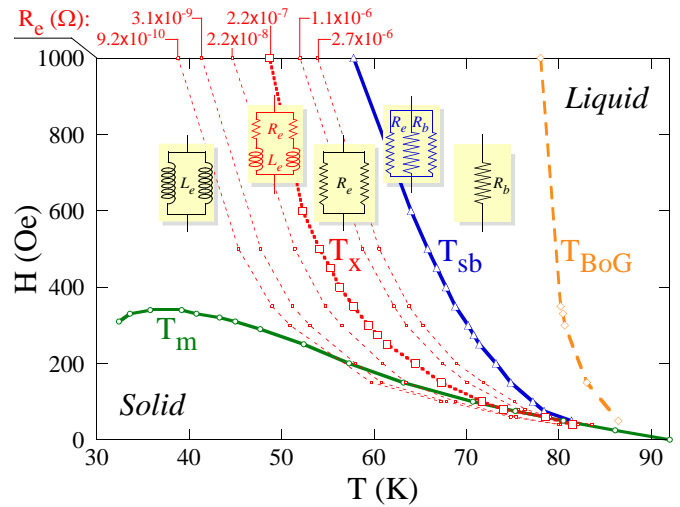}}
\caption{(color online) High temperature phase diagram of the vortex
matter in \bscco{} crystal B. $T_m$ is the first-order melting
($\circ$). Below $T_{sb}$ ($\triangle$) the edges shunt most of the
current from the bulk, while below the frequency dependent $T_x$
($\square$) the edge impedance becomes predominantly inductive. Ion
irradiated regions exhibit the Bose Glass transition,
$T_{\mathrm{BoG}}$ ($\diamond$), below which vortices are pinned to
columnar defects. The dotted lines represent equal edge-resistance
contours extracted from the frequency and field dependence of
$T_x$.} \label{fig1}
\end{figure}

The dynamic properties of the surface barriers have two types of
asymmetries. The first arises from the two edges of the sample
generally having somewhat different microscopic imperfections. It
leads to different edge resistances and therefore asymmetric current
distribution between the right and left edges. The $T_x$ line,
situated below $T_{sb}$ (Fig.~\ref{fig1}), marks the temperature
below which this right-left asymmetry sharply disappears and the
amplitudes of the two edge currents become equal
(Fig.~\ref{fig2}(f)). Hence, an antisymmetric current component
$\delta j^{(1)}$ (Fig.~\ref{fig2}(d)), that was present above $T_x$,
and its induced small symmetric contribution $\delta
B^{(1)}(x)=B^{(1)}(x,T_x+\frac{\delta T}{2})-B^{(1)}(x,T_x -
\frac{\delta T}{2})$ to the otherwise antisymmetric $B^{(1)}(x)$,
vanish below $T_x$.

The second type of asymmetry arises from the inherent asymmetry
between hard vortex entry and easy exit through the surface barriers
\cite{burlachkov.l.prb54}. Since the vortex entry and exit sides
swap during the ac cycle, this asymmetry gives rise to a unique
chiral second harmonic edge current, $\delta j^{(2)}$, and to the
corresponding second harmonic signal $\delta B^{(2)}(x)$
(Fig.~\ref{fig2}(e)). It builds up gradually upon cooling
(Fig.~\ref{fig2}(b)) until it pinches off sharply at $T_x$
(hereinafter determined at half of the $B^{(2)}(x)$ roll-off).
Accordingly, $T_x$ marks the temperature below which the vortex flow
turns insensitive to both the entry-exit and the right-left
asymmetries of the surface barriers.

The $T_x$ transition was previously ascribed to the advent of a new
vortex phase with a higher degree of order
\cite{fuchs.dt.prl80,sugano.r.pb284,nonomura.y.prl86,li.d.pc468,dietel.j.prb79}.
We show that $T_x$ has rather electrodynamic nature, arising from
the sample edges inductance. Following Ref.~\cite{brandt.eh.prb74}
we model the vortex dynamics by equivalent electric circuit with
three parallel channels - the bulk and two edges. Our measurements
show no dependence on the current magnitude (Fig.~\ref{fig2}(b),
dark vs. pale lines). Therefore, we follow the Ohmic model of
Ref.~\cite{brandt.eh.prb74} that assigns each channel geometrical
self and mutual inductances in series to their resistances.

\begin{figure} [!t]
\centering \mbox{\includegraphics[width=0.45\textwidth]{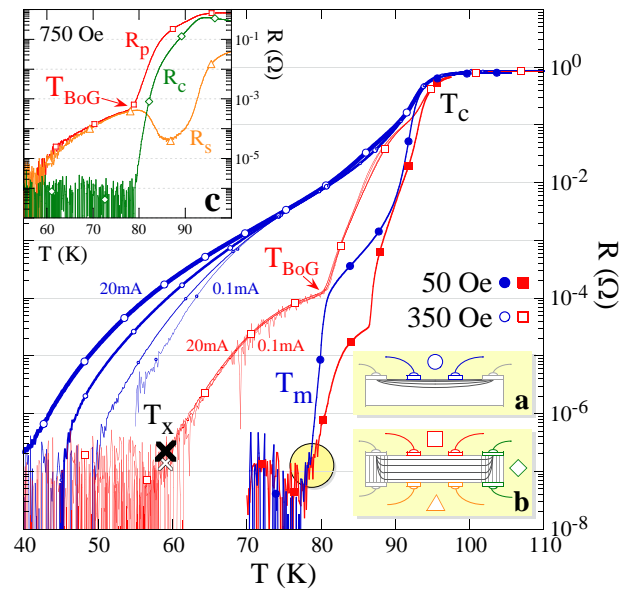}}
\caption{(color online) The pristine transport resistance (\bscco{}
crystal B) is non-Ohmic (shrinking $\circ$: $I_{ac}=$20, 5, 1, 0.1
mA, 73 Hz) due to non-uniform current flow along c-axis (inset a).
After contact irradiation ($\square$) and below $T_{\mathrm{BoG}}$
as the current penetrates uniformly (inset b) the resistance turns
Ohmic and matches that extracted from $T_x$ ($\times$). Below the
resistive step at $T_m$ the pristine ($\bullet$) and irradiated
($\blacksquare$) resistances are equal. Inset c: Multi-contact
measurement of a current-contact-irradiated sample (b). At low
fields and below $T_{\mathrm{BoG}}$, signaled by the vanishing
c-axis resistance ($\diamond$), current distributes uniformly across
the sample thickness, and the secondary resistance ($\vartriangle$)
equals the primary one ($\square$). } \label{fig3}
\end{figure}

The sample inductance is usually disregarded since it is
immeasurably small in direct transport measurements. Nevertheless,
it affects the current distribution in the sample. The effective
edge inductance, as opposed to the edge resistance, is temperature
independent and dictated solely by the geometry of the edges $L_e =
(\mu_0/4 \pi)l [\ln(2w/d)+1/4]$, where $l$, $w$, and $d$ are the
sample's length, width, and thickness, respectively
\cite{brandt.eh.prb74}. $T_x$ is the temperature below which the
edge impedance turns from being predominantly resistive to
inductive, $R_e(T_x)=2\pi f L_e$. Consequently, $T_{x}$ is frequency
dependent and allows sensitive determination of $R_e(T)$ similar to
the extraction of resistance from ac susceptibility
\cite{vdbeek.cj.prb43,steel.dg.prb45}. At 350 Oe and $f=73$ Hz we
measure $T_x\simeq 59$ K. Using $L_e=304$ pH, calculated from sample
geometry, we obtain $R_e(59$ K$)=1.4\times 10^{-7} \Omega$
(Fig.~\ref{fig3}, white cross). Note, that this value is 2-3 orders
of magnitude lower than the current dependent resistance measured
simultaneously in transport (open circles), which below $T_{sb}$
should reflect the edge resistance $R_e$.

The transport resistance measured in the geometry of Fig.~\ref{fig3}(a), however,
has a large contribution from the c-axis resistivity, $\rho_c$. Due to the extreme
anisotropy of \bscco{} $\rho_c$ is orders of magnitude larger than $\rho_{ab}$,
giving rise to nonlinearities and shear effects
\cite{busch.r.prl69,khaykovich.b.prb61}. The dissipation due to $\rho_c$ that
arises from current tunneling between the CuO$_2$ planes, however, is not accounted
in the electrodynamic considerations of the edge inductance \cite{brandt.eh.prb74}.
The non-uniformity of current flow along the $c$-axis can be remedied by
introducing columnar defects solely under the current contacts (see
Fig.~\ref{fig3}(b)). Below the Bose-glass transition, $T_{\mathrm{BoG}}$
(Fig.~\ref{fig1}, diamonds), the vortices become strongly pinned to the columnar
defects increasing the c-axis correlations and greatly reducing the sample
anisotropy \cite{nelson.dr.prl68,konczykowski.m.prb51,tamegai.t.jltp117}. This is
remarkably demonstrated in a multi-contact measurement of a sample irradiated in
such a manner (Fig.~\ref{fig3}(c)). Above $T_{\mathrm{BoG}}$ the high anisotropy
results in poor c-axis current penetration. Therefore, the primary resistance,
$R_p$, measured on the current injecting surface, is much higher than the secondary
resistance, $R_s$, measured on the opposite surface. At $T_{\mathrm{BoG}}$,
signaled by the plunging c-axis resistance $R_c$, the secondary resistance recovers
and equals the primary \cite{doyle.ra.prl77, seow.ws.prb53}. After current contact
irradiation the resistance, measured by voltage contacts in the central pristine
region of the sample (Fig.~\ref{fig3}, open squares), decreases by two orders of
magnitude, and turns Ohmic. Moreover, the resistive behavior now becomes fully
consistent with the inductive edge model. As shown below, all the data sets can be
fitted by a single parameter - an effective edge inductance of $L_e=$490 pH (black
cross in Fig.~\ref{fig3}). The similarity of this value to the calculated 304 pH is
remarkable, considering the crude modeling of the edges (round wires of diameter
$d$ \cite{brandt.eh.prb74}) and the uncertainties in various parameters.

\begin{figure} [!t]
\centering \mbox{\includegraphics[width=0.49\textwidth]{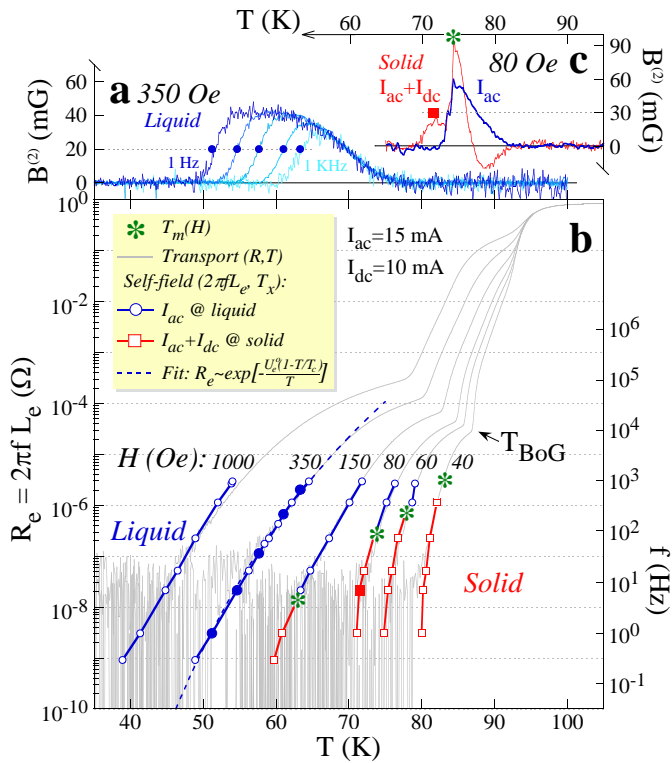}}
\caption{(color online) (a) Second-harmonic signals showing the
frequency dependence of $T_x$ ($\bullet$). (b) It enables to extract
the temperature dependent edge resistance ($\circ$) which is
thermally activated (dashed) and accurately matches the measured
transport resistance (thin lines). Below melting ($\ast$) a small dc
bias is required to extract the edge resistance ($\square$). (c)
This is seen in the second-harmonic signal (thick line) that
vanishes at melting ($\ast$) but recovers with bias (thin line)
before vanishing again at $T_x$ ($\blacksquare$).}\label{fig4}
\end{figure}

To further establish the role of the edge inductance in the $T_x$ transition we
extract $R_e(T,H)$ by repeatedly monitoring the current distribution with
frequencies ranging from 0.3 Hz to 1 kHz. At all frequencies $B^{(2)}$ rises
gradually with cooling (Fig.~\ref{fig4}(a)), as more current is shunted to the
edges, until it vanishes at a frequency-dependent temperature, $T_x(f)$. The
extracted edge resistance, $2\pi f L_e$ versus $T_x(f,H)$ (Fig.~\ref{fig4}(b),
circles), matches accurately the resistance measured in transport (thin lines), and
extrapolates it well below the transport noise floor. A fit to an Arrhenius
behavior (dotted line) yields an edge energy barrier $U_e^o\sim 18 T_c$.

The excellent agreement of the edge resistance extracted from
$T_x(f)$ to that measured in transport confirms that the edge
inductance drives the electrodynamic $T_x$ transition. In terms of
vortex dynamics, the resistive component of the edge impedance
arises from vortex dissipation thermally activated over the surface
barriers. Dissipation due to bulk vortex motion is negligible, since
bulk pinning is very weak, hence $R_b \gg R_e$. Nevertheless, bulk
redistribution of vortices reflects the inductive component of the
edge impedance. With changing current polarity during the ac cycle
the vortices, complying with the $B^{(1)}(x)$ profile of
Fig.~\ref{fig2}(f), shift from one side of the sample to the
opposite in association with an inductive $dB/dt$. At high
temperatures the number of vortices redistributing from right to
left during half a cycle due to $dB^{(1)}(x)/dt$ is still much
smaller than those that dissipatively cross the edges and move
across the sample. With cooling, however, the number of vortices
that cross the edges decreases exponentially.  Below $T_x$ it
becomes negligible compared to the number of redistributing
vortices. The second harmonic vanishes because vortex redistribution
in a `closed box' is necessarily antisymmetric. Our main finding is
that although the edge inductance is immeasurably small in transport
measurements, it completely governs vortex dynamics below $T_x$.

We now focus on the behavior of the edge resistance at melting. In
the pristine sample a sharp resistive drop \cite{fuchs.dt.prb54} is
observed at $T_m$ (Fig.~\ref{fig3}, solid dots), whereas after
contact irradiation the behavior is continuous (solid squares).
Moreover, the two curves merge at the lowest measurable resistance
(circled), indicating that in the presence of a uniform c-axis
current the edge resistance shows no sharp features. Hence, the
common resistive melting drop in \bscco{} arises from a sharp drop
in $\rho_c$.

Another intriguing feature at melting (Fig.~\ref{fig4}(c), asterisk)
is a concurrent vanishing of $B^{(2)}$ (thick line). A more detailed
analysis shows that this does not result from edge inductance, but
reflects an increased surface barriers in the vortex solid due to
enhanced c-axis correlations \cite{burlachkov.l.prb54}, which
enhances the current flow symmetry. To break this enhanced symmetry
below $T_m$ and probe the edge inductance we add a small dc current
bias. In its presence a finite second harmonic signal
(Fig.~\ref{fig4}(c), thin line) subsists below melting before
vanishing at $T_x$ (square). The detection of $T_x$ below melting
allows to extract the edge resistance within the solid phase
(Fig.~\ref{fig4}(b), squares), which fits perfectly and extrapolates
the transport resistance. Consequently, we find that the edge
resistance measured either in transport or by the edge inductance at
$T_x$ shows no pronounced feature at melting once the c-axis
contribution is removed by current contact irradiation. This
suggests that in highly anisotropic materials, such as \bscco{}, the
reported resistive melting step originates mostly from the loss of
c-axis correlations through a simultaneous sublimation into a
pancake gas \cite{glazman.li.prb43, fuchs.dt.prb55, colson.s.prl90}.
In contrast, the loss of inter-vortex correlations within the a-b
planes at melting has no significant mark in the vortex flow rate
through the edges.

Finally, our complete set of data is given by dotted lines in Fig.~\ref{fig1} that
represent contours of equal edge resistance (\emph{i.e.} measured at the same
frequency) spanning three and a half orders of magnitude in the $H-T$ phase
diagram. It is evident that the exponential temperature dependence of the edge
resistance becomes steeper on approaching the melting as the contour lines bunch
together. This is attributed to the enhanced stiffness of the pancake vortex
stacks. Nevertheless, we do not find a singular behavior at melting which would
have manifested itself in overlapping contour lines. On the contrary, once the
c-axis contribution is eliminated the edge resistance in solid joins smoothly that
of the liquid.

In summary, the spatial distribution of transport current and its
frequency dependence show that the $T_x$ line, tentatively ascribed
to a phase transition of vortex matter in \bscco{}, is rather an
electrodynamic transition. Consequently, in the wide field and
temperature range that lies below the $T_x$ line the inductance of
the sample edges, usually dismissed in transport experiments,
dominates the current flow and causes ac displacement of vortices
across the bulk without crossing the sample edges. At $T_x$, the
inductive part of the edge impedance equals the resistive part,
which allows measurement of resistance down to two orders of
magnitude below the transport noise. In the vicinity of the melting
transition, we find that once the c-axis contribution is eliminated
via ion irradiation of the current contacts, the edge resistance
shows no sharp features at melting.

\begin{acknowledgements}
We thank H.~Shtrikman for GaAs heterostructures and M.~Konczykowski for ion
irradiation. This work was supported by the German-Israeli Foundation (GIF). HB
acknowledges the support of the Adams Fellowship Program of the Israel Academy of
Sciences and Humanities and EZ the US-Israel Binational Science Foundation (BSF).
\end{acknowledgements}


\begin{thebibliography}{10}

\bibitem{blatter.g.rmp66}
G.~Blatter \emph{et al.},
\newblock {\em Rev. Mod. Phys.}{ \bf 66}, 1125 (1994).

\bibitem{mikitik.gp.prb68}
G.~P.~Mikitik and E.~H.~Brandt,
\newblock {\em Phys. Rev. B}{ \bf 68}, 054509 (2003).

\bibitem{zeldov.e.nat375}
E.~Zeldov \emph{et al.},
\newblock {\em Nature}{ \bf 375}, 373 (1995).

\bibitem{schiling.a.prl78}
A.~Schilling \emph{et al.},
\newblock {\em Phys. Rev. Lett.}{ \bf 78}, 4833 (1997).

\bibitem{safar.h.prl68}
H.~Safar \emph{et al.},
\newblock {\em Phys. Rev. Lett.}{ \bf 68}, 2672 (1992).

\bibitem{vdbeek.cj.pc195}
C.~J.~van der Beek \emph{et al.},
\newblock {\em Physica C}{ \bf 195}, 307 (1992).

\bibitem{beidenkopf.h.prl95}
H.~Beidenkopf \emph{et al.},
\newblock {\em Phys. Rev. Lett.}{ \bf 95}, 257004 (2005).

\bibitem{beidenkopf.h.prl98}
H.~Beidenkopf \emph{et al.},
\newblock {\em Phys. Rev. Lett.}{ \bf 98}, 167004 (2007).

\bibitem{fuchs.dt.prl80}
D.~T.~Fuchs \emph{et al.},
\newblock {\em Phys. Rev. Lett.}{ \bf 80}, 4971 (1998).

\bibitem{forgan.em.czec46}
E.~Forgan,
\newblock {\em Czech. J. Phys}{ \bf 46}, 1571 (1996).

\bibitem{shibauchi.t.prl83}
T.~Shibauchi \emph{et al.},
\newblock {\em Phys. Rev. Lett.}{ \bf 83}, 1010 (1999).

\bibitem{ando.y.prb59}
Y.~Ando and K.~Nakamura,
\newblock {\em Phys. Rev. B}{ \bf 59}, R11661 (1999).

\bibitem{kimura.k.pb284}
K.~Kimura \emph{et al.},
\newblock {\em Physica B}{ \bf 284}, 717 (2000).

\bibitem{eltsev.y.pc341}
Y.~Eltsev \emph{et al.},
\newblock {\em Physica C}{ \bf 341}, 1107 (2000).

\bibitem{sugano.r.pb284}
R.~Sugano \emph{et al.},
\newblock {\em Physica B}{ \bf 284}, 803 (2000).

\bibitem{nonomura.y.prl86}
Y.~Nonomura and X.~Hu,
\newblock {\em Phys. Rev. Lett.}{ \bf 86}, 5140 (2001).

\bibitem{li.d.pc468}
D.~Li, P.~Lin and B.~Rosenstein,
\newblock {\em Physica C}{ \bf 468}, 1245 (2008).

\bibitem{dietel.j.prb79}
J.~Dietel and H.~Kleinert,
\newblock {\em Phys. Rev. B}{ \bf 79}, 014512 (2009).

\bibitem{samoilov.av.prl76}
A.~V.~Samoilov \emph{et al.},
\newblock {\em Phys. Rev. Lett.}{ \bf 76}, 2798 (1996).

\bibitem{glazman.li.prb43}
L.~I.~Glazman and A.~E.~Koshelev,
\newblock {\em Phys. Rev. B}{ \bf 43}, 2835 (1991).

\bibitem{horovitz.b.prl80}
B.~Horovitz and T.~R.~Goldin,
\newblock {\em Phys. Rev. Lett.}{ \bf 80}, 1734 (1998).

\bibitem{carruzzo.hm.pmb77}
H.~M.~Carruzzo and C.~C.~Yu,
\newblock {\em Philos. Mag. B}{ \bf 77}, 1001 (1998).

\bibitem{frey.e.prb49}
E.~Frey, D.~R.~Nelson and D.~S.~Fisher,
\newblock {\em Phys. Rev. B}{ \bf 49}, 9723 (1994).

\bibitem{feigelman.mv.pc167}
M.~Feigel'man, V.~Geshkenbein and A.~Larkin,
\newblock {\em Physica C}{ \bf 167}, 177 (1990).

\bibitem{bean.cp.prl12}
C.~P.~Bean and J.~D.~Livingston,
\newblock {\em Phys. Rev. Lett.}{ \bf 12}, 14 (1964).

\bibitem{zeldov.e.prl73}
E.~Zeldov \emph{et al.},
\newblock {\em Phys. Rev. Lett.}{ \bf 73}, 1428 (1994).

\bibitem{fuchs.dt.nat391}
D.~T.~Fuchs \emph{et al.},
\newblock {\em Nature}{ \bf 391}, 373 (1998).


\bibitem{burlachkov.l.prb54}
L.~Burlachkov, A.~E.~Koshelev and V.~M.~Vinokur,
\newblock {\em Phys. Rev. B}{ \bf 54}, 6750 (1996).

\bibitem{brandt.eh.prb74}
E.~H.~Brandt, G.~P.~Mikitik and E.~Zeldov,
\newblock {\em Phys. Rev. B}{ \bf 74}, 094506 (2006).

\bibitem{vdbeek.cj.prb43}
C.~J.~van der Beek and P.~H.~Kes,
\newblock {\em Phys. Rev. B}{ \bf 43}, 13032 (1991).

\bibitem{steel.dg.prb45}
D.~G.~Steel and J.~M.~Graybeal,
\newblock {\em Phys. Rev. B}{ \bf 45}, 12643 (1992).

\bibitem{busch.r.prl69}
R.~Busch \emph{et al.},
\newblock {\em Phys. Rev. Lett.}{ \bf 69}, 522 (1992).

\bibitem{khaykovich.b.prb61}
B.~Khaykovich \emph{et al.},
\newblock {\em Phys. Rev. B}{ \bf 61}, R9261 (2000).

\bibitem{nelson.dr.prl68}
D.~R.~Nelson and V.~M.~Vinokur,
\newblock {\em Phys. Rev. Lett.}{ \bf 68}, 2398 (1992).

\bibitem{konczykowski.m.prb51}
M.~Konczykowski \emph{et al.},
\newblock {\em Phys. Rev. B}{ \bf 51}, 3957 (1995).

\bibitem{tamegai.t.jltp117}
T.~Tamegai \emph{et al.},
\newblock {\em J. Low Temp. Phys.}{ \bf 117}, 1363 (1999).

\bibitem{doyle.ra.prl77}
R.~A.~Doyle \emph{et al.},
\newblock {\em Phys. Rev. Lett.}{ \bf 77}, 1155 (1996).

\bibitem{seow.ws.prb53}
W.~S.~Seow \emph{et al.},
\newblock {\em Phys. Rev. B}{ \bf 53}, 14611 (1996).

\bibitem{fuchs.dt.prb54}
D.~T.~Fuchs \emph{et al.},
\newblock {\em Phys. Rev. B}{ \bf 54}, R796 (1996).

\bibitem{fuchs.dt.prb55}
D.~T.~Fuchs \emph{et al.},
\newblock {\em Phys. Rev. B}{ \bf 55}, R6156 (1997).

\bibitem{colson.s.prl90}
S.~Colson \emph{et al.},
\newblock{\em Phys. Rev. Lett.}{ \bf 90}, 137002 (2003).
\end{thebibliography}
\end{document}